\documentclass[seceq,supplement]{ptptex}

\title{
Gauge glass in two dimensions%
}

\author{
Lei-Han \textsc{Tang}%
}

\inst{
Department of Physics, Hong Kong Baptist University, Kowloon Tong, Hong Kong SAR, China
}

\abst{
The gauge glass model offers an interesting example of a randomly frustrated system
with a continuous O(2) symmetry. In two dimensions, the existence of a glass phase
at low temperatures has long been disputed among numerical studies. 
To resolve this controversy, we examine the behavior of vortices whose 
movement generates phase slips that destroy phase rigidity at large distances.
Detailed analytical and numerical studies of the corresponding
Coulomb gas problem in a random potential establish that
the ground state, with a finite density of vortices, is polarizable with a scale-dependent
dielectric susceptibility. Screening by vortex/antivortex pairs of arbitrarily large size
is present to eliminate the logarithmic divergence of the Coulomb energy of a single
vortex. The observed power-law decay of the Coulomb interaction between vortices with 
distance in the ground state leads to a power-law divergence of the glass correlation
length with temperature $T$. It is argued that free vortices possess a 
bound excitation energy and a nonzero diffusion constant at any $T>0$.                     
 }

\begin{document}

\maketitle

\section{Introduction}

In a seminal review that summarized kinetic and thermodynamic properties of 
glass forming substances, Austen Angell\cite{Angell} made the remark that 
``glass, in the popular and basically correct conception, is a liquid that has lost 
its ability to flow''. 
In other words, structurally glass is indistinguishable from liquid yet 
mechanically it behaves much like solid. The key issue to understand the glass 
state thus lies in how a disordered structure can withstand shear, which is a form of 
``rigidity'' that is usually attributed to crystalline solid but not liquid\cite{Anderson}. 

A second, also universal property of glass, is the linear specific
heat curve at low temperatures\cite{Phillips}. Such temperature dependence is seen 
in metals but is usually absent in insulating materials. 
Soon after the pioneering discovery of this behavior by Zeller and Pohl\cite{Zeller}
in 1971, Anderson, Halperin and Varma\cite{Anderson71} proposed a theoretical
explanation by assuming the existence of two-level ``tunneling states'' whose energy gap
has a continuous distribution across zero. A mechanism similar to the electron-hole 
pair excitations was invoked to explain the observed thermodynamic behavior.

Given these two common and unique features of a glass state, one can not help but
wonder whether there is an intricate interplay between them. 
The linear specific heat implies existence of nearly degenerate 
local states which can plausibly arise from the many different 
local atomic arrangements in glass. 
Rigidity against shear, on the other hand, is a large distance behavior. In the case of 
crystalline solid, it hinges on the suppression or pinning of dislocations which are 
topological defects in a periodic structure. 
In contrast, glass does not have a definitive local order and hence
identification of dislocation-like defects can be quite ambiguous. Nevertheless, 
the fact that glass does not flow suggests that response of the system to shear stress,
at least on time scales of interest, can be described by atomic displacements 
that vary only slowly in space as in a crystalline solid. It is then interesting to
ask:  To what extent the phenomenology of elastic deformations and  
dislocations for crystalline solid can be carried over to describe 
glass, despite the existence of a large number of 
alternative and thermally accessible local atomic configurations?

The XY model on a square lattice has played a key role in the theory of melting
of two-dimensional (2D) solid\cite{Nelson}. In the same vein, one may argue that
valuable lessons can be learned about glass by studying a disordered version of the XY model.
A simple way to introduce disorder without destroying the global O(2) symmetry
is to assign preferred but random phase shifts to bonds connecting neighboring sites.
The ground state of the system is then frustrated and, for sufficiently strong disorder,
loses the long-range phase alignment in the unfrustrated case due to the excitation of free vortices. 
Hence we have a model system that
contains some of the essential features of real glass. One great advantage of the 
lattice model is that defects that play the role of dislocations in a solid can be 
unambiguously identified. This way, their energetics can be studied in detail using well
established analytic and numerical methods.

The gauge glass model, originally proposed in the study of disordered superconductors
in a magnetic field\cite{Ebner,FTY}, is an extreme form of the randomly frustrated XY model.
Previous simulational studies of the model in two dimensions have not reached consensus 
on the existence of a low-temperature glass phase\cite{FTY,RY93,Akino,Katzgraber,NW04,CP99,Kim,HMK}.
The spin-glass susceptibility $\chi_{\rm SG}$, which 
measures the degree of spatial correlations in the system,
is found to increase sharply as temperature is lowered. 
However, due to the relatively small system sizes examined and the need
to sample a large number of disorder configurations, 
it is difficult to make definitive conclusions about the behavior
of $\chi_{\rm SG}(T)$ in the thermodynamic limit based on these studies.

An alternative method\cite{tang05} is to examine 
more closely the behavior of microscopic degrees of freedom, in this case ``spin-wave'' 
and vortex excitations over the ground state, to determine the system's thermal properties 
and large-distance behavior. Detailed numerical and analytic investigations
suggest that the typical energy $\Delta E$ associated with phase deformations
on length scale $R$ scales as $\Delta E\sim R^\theta$, where $\theta\simeq -0.45$ is a 
negative exponent\cite{FTY,RY93,Akino,Katzgraber,NW04,tang05}. 
Equating $\Delta E$ with the thermal energy $k_BT$ (henceforth we set the Boltzmann
constant $k_B=1$), we identify a thermal correlation length
$\xi_{\rm th}(T)\sim T^{-1/|\theta|}$ at temperature $T$. 
In the infinite size limit, the spin-glass susceptibility varies with temperature
as $\chi_{\rm SG}\sim\xi_{\rm th}^{2-\eta}\sim T^{-(2-\eta)/|\theta|}$, where $\eta$ is an
exponent due to spin-wave fluctuations which is generally small and vanishes
as $T\rightarrow 0$. Therefore $T=0$ is a critical point with regard to
long-range glass ordering in the 2D gauge glass.

In what follows we recapture the main steps of reasoning and the supporting numerical 
results that led to the above conclusion. Dynamic effects that could possibly complicate 
a direct application of the equilibrium considerations to relevant experiments are
discussed briefly at the end.

\section{The Model and Vortex Hamiltonian}

The two-dimensional randomly frustrated XY model is defined by the Hamiltonian:
\begin{equation}
H=-J\sum_{\langle ij\rangle}\cos(\phi_i-\phi_j-A_{ij}).
\label{gauge-glass-H}
\end{equation}
Here $\phi_i$ is the phase variable on site $i$ of a two-dimensional square lattice, 
$J$ is the coupling constant, and $A_{ij}$ are independently distributed, quenched random 
variables with zero mean and variance $\sigma$. Summation is over nearest
neighbor pairs of sites. The gauge glass corresponds to the extreme
form where the $A_{ij}$'s are uniformly distributed on $[-\pi,\pi)$.
The model possesses O(2) symmetry so that an overall
shift of the phase $\phi_i\rightarrow\phi_i+c$ does not change the
system energy $H$. This gives rise to the gapless spin-wave excitations,
which behave similarly as phonons in a particle system. 
The second class of excitations are vortices, which can be 
identified as the rotational component of the lattice phase gradient:
\begin{equation} 
\nabla_{\rm lattice}\phi({\bf r})=\nabla\phi^{\rm sw}({\bf r})
+\sum_k m_k\hat{z}\times{{\bf r}-{\bf r}_k\over|{\bf r}-{\bf r}_k|^2}.
\label{decomposition}
\end{equation}
Here $m_k=0,\pm 1$ denotes the vortex charge on the plaquette $k$ at ${\bf r}_k$, and
$\hat{z}$ is a unit vector normal to the plane. 

In the ``pure'' case $A_{ij}=0$, Eq. (\ref{gauge-glass-H}) reduces to the well-known
XY model, which has a finite temperature Kosterlitz-Thouless (KT) transition at 
$T_{\rm KT}\simeq 0.89J$.
When $T<T_{\rm KT}$, vortices ($m_k>0$) and antivortices ($m_k<0$) form bound pairs.
The population of these thermally excited pairs behave as a polarizable medium and
reduces the effective spin-wave stiffness (or rigidity) $J_{\rm eff}<J$ at large distances.
The KT transition temperature is set by $T_{\rm KT}/J_{\rm eff}=\pi/2$. 
For $T>T_{\rm KT}$, one is in a disordered phase where proliferation of vortex/antivortex
pairs of large size reduces vortex interaction to a finite range. Phase rigidity on sufficiently 
large length scales is lost.

The random phase shifts $A_{ij}$ contribute to the energy of a vortex through a modification of the
vortex core energy $E_c$ as well as interaction with the elastic distortions introduced by the 
vortex. These effects can be described quantitatively by rewriting
the energy function (\ref{gauge-glass-H}) in the form\cite{rubi83}
$H=H_{\rm sw}(\{\phi^{\rm sw}\})+H_{\rm v}(\{m_k\})$, where
\begin{equation}
H_{\rm v}(\{m_k\})=\sum_k[m_k^2E_c+m_kV({\bf r}_k)]-2\pi J\sum_{k<k'}m_{k} m_{k'}
\ln {r_{kk'}\over a}
\label{coulomb-gas}
\end{equation}
is the Coulomb gas Hamiltonian.
Here $a$ is the lattice constant, and $r_{kk'}$ is the distance between sites 
$k$ and $k'$. The random potential $V({\bf r})$ is given by a sum of dipolar
potentials produced by random phase shifts $A_{ij}$. 
Its statistics is specified by a logarithmically growing variance 
$\langle V^2({\bf r})\rangle\simeq 2\pi\sigma J^2\ln (L/a)$ with linear system size $L$, and 
logarithmic spatial correlations 
$\langle [V({\bf r})-V({\bf r}')]^2\rangle\simeq 4\pi\sigma J^2\ln(|{\bf r}-{\bf r}'|/a)$.

Much can be learned about properties of the randomly frustrated XY model and
the corresponding Coulomb gas from the statistical mechanics of
a single vortex in a random potential $V({\bf r})$.\cite{cha95,nskl}
Through a mapping to the directed polymer on Cayley tree problem\cite{tang96},
the free energy of the vortex, averaged
over the disorder $A_{ij}$, can be computed analytically for large $L$,
\begin{equation}
\langle f\rangle=\left\{
\begin{array}{l}
E_c+\pi J\Bigl(1-{2T\over\pi J}-{\sigma J\over T}\Bigr)\ln{L\over a},\;\;\mbox{for}\; T>T_g;\\
\\
E_c+\pi J\Bigl(1-\sqrt{8\sigma\over\pi}\Bigr)\ln{L\over a},\;\;\mbox{for}\; T<T_g.
\end{array}\right.
\label{single_v}
\end{equation} 
Here $T_g=J\sqrt{\pi\sigma/2}$ is the glass transition temperature of the single vortex problem. 
Note that for $T<T_g$, the entropy of the vortex is essentially zero. In this case,
the equilibrium state is dominated by the site of the lowest $V({\bf r})$ in the system. 
Delocalization of the vortex takes place at $T=T_g$.
Setting $\langle f\rangle =0$, one obtains the boundary or stability limit of the ordered phase 
in the randomly frustrated XY model against single vortex excitations.
On the $T$-$\sigma$ plane, the ordered phase ($\langle f\rangle>0$) is bordered by
the inverted half parabola $\sigma=T/J-(2/\pi)(T/J)^2$ for $\pi/4<T/J<\pi/2$, and the
flat line $\sigma=\sigma_c=\pi/8$ for $T/J<\pi/4$. The effect of other vortex-antivortex
pairs on (\ref{single_v}) can be calculated in a renormalization group (RG) scheme
which yields detailed properties of the finite temperature KT-like transition in the
randomly frustrated XY model when the disorder strength is less than $\sigma_c$.
\cite{nskl,tang96,carpentier}

The gauge glass model corresponds to a $\sigma=\pi^2/3\simeq 8.4\sigma_c$, way above
the stability limit of the vortex-free phase.
Its ground state contains a finite density of vortices whose spatial distribution depends
both on the random potential $V({\bf r})$ and on their mutual Coulomb interaction. 
Understanding various properties of
this seemingly random but frozen structure (as compared to the thermally disordered state
above the KT transition) and possible correlations within is the focus of our 
discussion below.

\section{The Ground State}

Finding the ground state of the vortex Hamiltonian (\ref{coulomb-gas}) for an arbitrary
random potential $V({\bf r})$ is a computationally hard problem. For small system sizes,
our experience shows that this task can be accomplished by a greedy algorithm\cite{tang05}. 
The ground state search process yields important 
insights about the organization of low energy states.

In a single run of the greedy algorithm, a random configuration of vortices 
and antivortices in equal numbers is generated as the initial state. 
The total potential $\tilde V({\bf r})$ of a vortex
at site ${\bf r}$, defined as the sum of $V({\bf r})$ and the Coulomb potentials from 
other vortices in the system, is monitored during subsequent updates which follow
a downward path in total system energy. Each updating event consists of addition
or removal of a single vortex/antivortex pair, provided the vortex charge on any given
site does not exceed $+1$ or fall below $-1$.
The locations ${\bf r}_+$ and ${\bf r}_-$ of the newly introduced vortex and antivortex
are selected to minimize the pair energy 
$E_{\rm pair}=E_0+\tilde V({\bf r}_+)-\tilde V({\bf r}_-)+2\pi J\ln(|{\bf r}_+-{\bf r}_-|/a)$,
where $E_0=2E_c$ if the two sites ${\bf r}_+$ and ${\bf r}_-$ were previously empty,
$E_0=0$ if one of the two sites was occupied by a vortex of opposite sign, and $E_0=-2E_c$ 
if both sites were occupied by vortices of opposite charge. 
The last two cases actually correspond to 
moving a vortex (antivortex) to a new position, or removal of a vortex/antivortex pair,
respectively. The process stops when the system energy can no longer be decreased
through such moves. In an improved version, we allow the update to continue even when
the lowest $E_{\rm pair}$ (among all possible pair positions) is positive, but stops when a previously 
identified minimum energy state (in the same run) is revisited. This additional step often 
yields a better minimum energy state for larger systems. 

To ensure that the minimum energy state so identified is 
indeed the ground state for a given $V({\bf r})$, repeated runs from different initial 
conditions are performed. A sequence of minimal energy states is obtained in consecutive runs.
The typical intervals between successive visits to the lowest energy state {\it a posteriori} are monitored
to estimate the number of runs needed to reach the ground state during random sampling
of initial conditions. For systems of linear size $L=16$ or smaller, 100 runs per
disorder configuration is usually sufficient.

In addition to the ground state, the above scheme also yields a number of low energy states
at the end of each minimization run. We have compared the vortex 
positions of these states, and found that those with energies close to the ground state 
differ from one another only at a few vortex positions\cite{tang05}. 
Thus the ground state of the system is fairly unique despite its disordered nature.
The fact that excited states contain only a low density of excess vortex/antivortex
pairs suggests a perturbative scheme to capture the low temperature properties
of the system.

Before outlining such a calculation in the next section, we mention some important
properties of the total vortex potential $\tilde V({\bf r})$ as seen in the numerical
experiments. The ``bare'' potential
$V({\bf r})$ generated by the random dipoles has an interesting spatial structure\cite{tang96}. 
It can be shown that the Fourier components
of $V({\bf r})$ are statistically independent of each other.
Interestingly, during each minimization run of the greedy algorithm,
variance of the total potential $\tilde V({\bf r})$ continues to 
decrease and, at the end of the process, reaches a value much smaller than the 
variance of the bare $V({\bf r})$.
Fourier analysis of the final $\tilde V({\bf r})$ shows that the reduction
is much stronger at smaller values of the wavevector ${\bf k}$. Thus, while the
newly introduced vortex/antivortex pairs take the lowest energy positions during
minimization, they act simultaneously as screening dipoles to reduce the
``electric field'' generated by the quenched random dipoles and other vortex charges
already in the system.
Numerical results\cite{tang05} suggest that the reduction factor of
the variance of $\tilde V({\bf k})$ scales approximately linearly with the wavenumber 
$|{\bf k}|$ in the ground state. These observations motivate the phenomenological RG treatment
below.

\section{The Renormalization Group Analysis}

The Coulomb gas problem without disorder $V({\bf r})$ was analyzed in the classic
work by Kosterlitz and Thouless following a real-space RG scheme\cite{kost73,kost74}.
The basic idea of the RG treatment is to describe a charge-neutral Coulomb gas
as weakly interacting vortex/antivortex pairs of varying size, and characterize the effect 
of smaller pairs on larger ones in terms of dielectric screening.
Each vortex/antivortex pair has a dipole moment ${\bf p}={\bf r}_+-{\bf r}_-$
and hence its energy changes by an amount $\Delta E_{\rm pair}=-{\bf E}\cdot{\bf p}$ in
the presence of an applied external field ${\bf E}$, or for that matter, respond to 
the field generated by distant vortices. 
Thermally excited vortex/antivortex pairs do not have a preferred orientation
of their own. Their polarizability is derived from the energy gain when ${\bf p}$
is oriented in the direction of {\bf E}. The polarization in turn reduces
the strength of the field {\bf E} at large distances.

To turn the above idea into a RG procedure, one introduces
a running cutoff size $R$ of vortex/antivortex pairs. Each time only pairs in
a particular size range is considered, starting from the smallest size set
by the lattice constant. Dielectric screening of the field at large distances is computed
analytically following a linear response theory at small vortex/antivortex pair densities.
The result of this calculation is a two-parameter RG flow that includes i) the
renormalized vortex core energy $E_c(R)$ (or more precisely the pair chemical potential)
that controls the pair density on scale $R$,
and ii) the running coupling constant $J(R)$ after taking into account the screening
effect from vortex/antivortex pairs of size less than $R$. 
Since the vortex/antivortex pair density vanishes
in the limit $R\rightarrow\infty$ in the ordered phase of the XY model
(including the phase boundary), the perturbative RG equations give an asymptotically
exact description of various scaling properties of the transition.\cite{kost73,kost74}

The disorder potential $V({\bf r})$ introduces spatial variations of the vortex/antivortex
pair energy which may even go negative at certain locations. The latter situation
gives rise to ground state vortices whose polarizability needs a separate consideration
from those of the thermally excited ones\cite{tang96}. For $\sigma<\sigma_c$, the average number of
such vortex/antivortex pairs in an area of the pair size $R$ goes to zero as
$R\rightarrow\infty$. The main effect of the disorder in this case is to reduce
the effective vortex core energy $E_c$, thereby shifting the transition temperature
to a lower value $T_c(\sigma)<T_{\rm KT}$. Renormalization effects due to both thermal and
disorder-induced vortices have been analyzed in detail\cite{nskl,tang96}.
Essential features of the transition remains the same as in the pure case
but certain details are modified.

Vortex/antivortex pairs induced by the disorder potential $V({\bf r})$ are
strongly localized and hence do not respond to a weak external 
field via a continuous change of their orientation. The polarizability of such
pairs is a statistical effect at the population level. Only pairs with 
energies close to zero participate in the process. For a pair whose energy is just
below zero, the extra energy $\Delta E_{\rm pair}$ may render the total energy positive
if the dipole moment ${\bf p}$ is against the applied field {\bf E}. 
Consequently, the pair should disappear from the ground state in the presence
of {\bf E}. On the other hand, a pair whose energy is just above zero may
become favorable under an applied field. Both cases contribute to a field-induced
polarization ${\bf P}=\chi {\bf E}$ where the dielectric susceptibility
$\chi$ is proportional to the density of pair states $\rho(0)$ at zero energy. 
Separating pairs into logarithmically binned sizes, the contribution to $\chi$
from those in the size range $R$ to $R+dR$ can be written as
$\chi_R=R^2\rho_R(0)\times (dR/R)$, where the subscript $R$ indicates pairs
in the size range. Writing $dl=dR/R$, reduction of the coupling constant
for distant vortices due to screening by this group of pairs takes the form,
\begin{equation}
dJ^{-1}/dl=4\pi^2\hat\rho(0),
\label{RG-J}
\end{equation}
where $\hat\rho(0)=R^2\rho_R(0)$ is the density of pair states at zero energy
in an area of pair size. (Refer to Ref. 16 for a more precise definition of the
quantities involved and a derivation of this equation.)

Equation (\ref{RG-J}) is generally valid for the ground state of the
classical Coulomb gas which is polarizable despite the localization effects.
For $\sigma<\sigma_c$, the gain in energy due to the disorder potential $V({\bf r})$
is usually not sufficient to bring the energy of a vortex/antivortex pair to below 
zero so that, in the majority of cases, the number of pairs in any given
region of pair size $R$ goes to zero as $R\rightarrow\infty$.
A very different situation is encountered in the gauge glass, where 
$\hat\rho(0)$ is finite on all scales. In fact, since there is no energy
gap for vortex/antivortex pair excitation, the only remaining energy
scale for pair states on length scale $R$ is $J(R)$. The density of pair states
at zero energy is expected to be inversely proportional to $J(R)$,
\begin{equation}
\hat\rho(0)=cJ^{-1}.
\label{rho_0}
\end{equation}
Integration of Eq. (\ref{RG-J}) under the assumption (\ref{rho_0}) yields
a power-law decaying coupling constant, $J(R)\simeq J_BR^\theta$, where
$\theta=-4\pi^2c$ is a negative exponent.

The above description of the gauge glass ground state as a polarizable medium matches
well with observations from the numerical experiments presented in the
previous section. Once the seemingly complex vortex configuration is described
in terms of paired vortices, the dominant interaction of a vortex with the disorder 
and with other vortices in the system
is captured by the effective applied field $\tilde V({\bf r})$. 
In the ground state and at low temperatures, only pair states with energies
sufficiently close to the ``Fermi energy'' $\epsilon_F=0$ participate in the
screening process. When considering a pair of large size, each of the two constituent
vortices generates a local field that polarizes smaller pairs in the neighborhood.
Therefore large pairs are ``dressed'' by smaller ones much like electrons in
Landau's Fermi liquid theory.
The greedy algorithm introduced above looks for vortex pair configurations
based on $\tilde V({\bf r})$ alone, which is effective when the system
size is small, but its performance deteriorates on larger scales as the selection
process should also include the screening pairs.

\section{Monte Carlo Simulation at Finite Temperatures}

Both the numerical and the RG analysis of the Coulomb gas ground state
suggest diminishing energy scale with growing length scale for vortex/antivortex
pair excitations. The zero temperature state is thus a critical point.
At low but finite temperatures, the system exhibits rigidity 
(i.e., a finite spin-wave stiffness) on length scales below 
$\xi_{\rm th}(T)\sim T^{-1/|\theta|}$, but is disordered due to thermal excitation
of vortices on larger scales.

A quantitative measure of the glass order in the XY model is the
two-point correlation function
\begin{equation}
C_{\rm SG}(r_{ij})=\langle|\overline{e^{i(\phi_i-\phi_j)}}|^2\rangle,
\label{correlation}
\end{equation}
where the overline bar denotes thermal average, and $\langle\cdot\rangle$ denotes
average over the disorder realizations. Both spin-wave and vortex excitations
contribute to the decrease of $C_{\rm SG}(r)$ with distance $r$.
For $r<\xi_{\rm th}(T)$, $C_{\rm SG}(r)\sim r^{-\eta}$ due to spin-wave fluctuations,
where $\eta=T/(2\pi J(r))$.
This behavior is expected to crossover to exponential decay for $r>\xi_{\rm th}(T)$
due to free vortex excitations. Another way to characterize possible glass
order is to consider the overlap function
\begin{equation}
q_{ab}={1\over N}\Bigl|\sum_{j=1}^Ne^{i(\phi_j^a-\phi_j^b)}\Bigr|
\label{overlap}
\end{equation}
of two spin configurations $\phi_j^a$ and $\phi_j^b$. Here $N=L^2$ is the total
number of sites on the lattice.
In the equilibrium ensemble, the overlap distribution is
given by $P(q)=\langle\overline{\delta(q-q_{ab})}\rangle$, 
where the thermal average is taken over two independent replicas 
$a$ and $b$ under the same disorder. 
Introducing the Fourier transform
$\chi_{\rm SG}(T,{\bf k})=\sum_jC_{\rm SG}(r_{ij})e^{i{\bf k}\cdot {\bf r}_{ij}}$,
the spin-glass susceptibility is expressed as
\begin{equation}
\chi_{\rm SG}(T)\equiv\chi_{\rm SG}(T,{\bf k}=0)=N\int q^2P(q)dq.
\label{chi}
\end{equation}
In the disordered phase, the overlap $q_{ab}$ between two typical equilibrium
configurations goes down as $N^{-1/2}$ in the large $N$ limit.
In this case, the distribution $P(q)$ has a single peak at $q=0$ and
a width given by $(\chi_{\rm SG}/N)^{1/2}$. On the other hand, a finite weight
of $P(q)$ at nonzero $q$ values in the thermodynamic limit yields a diverging 
$\chi_{\rm SG}$.

It is clear from Eq. (\ref{chi}) that computation of the spin-glass
susceptibility $\chi_{\rm SG}$ requires full information of the overlap
distribution $P(q)$. This condition imposes strong equilibration 
requirements on Monte Carlo methods used to calculate $\chi_{\rm SG}$.
In the usual Metropolis importance sampling scheme at a constant temperature,
a sequence of configurations are generated through phase displacement
at randomly selected sites. This type of Monte Carlo move is quite inefficient in
achieving large distance vortex movement, particularly when such movement
requires crossing over energy barriers along a continuous path.
To overcome this problem, we adopted a modified version of Berg's
entropic sampling scheme\cite{berg,tang_MCMC}. The fact that the system 
shuttles between high energy and low energy states in a single Monte Carlo 
run allows much better coverage of the low energy states.
In addition, when the simulation is carried out on a PC cluster with
a large number of nodes, each node performs an independent walk
in configuration space, which allows efficient computation of $P(q)$
by crossing configurations from different nodes.

With the help of the entropic sampling scheme, we were able to equilibrate
systems of size up to $48\times 48$ down to $T=0.04J$. The main results of
the simulation are reported in Ref. 16.
The specific heat as a function of temperature indeed has a linear component as
$T\rightarrow 0$. This behavior can be attributed to the vortex/antivortex
pair excitations which are gapless in the gauge glass. No indication of
a finite-temperature transition is observed from the specific heat curve.
Finite-size scaling analysis is performed on the spin-glass susceptibility data.
After taking into account spin-wave contributions, excellent data collapse
is achieved using the scaled temperature $TL^{|\theta|}$ for systems ranging
from $L=6$ to $L=48$, reconfirming $T=0$ to be the critical point with respect
to glassy order.

\section{Discussion and Conclusions}

Returning to the two universal properties of a glass mentioned in 
Sec. 1, we see that the continuous energy spectrum of alternative local spin
configurations (which are local minima of the energy function) 
in the 2D gauge glass model indeed gives rise to a linear
specific heat as in general two-level systems. The number of such states
is extensive. The Coulomb gas offers a concrete representation of
how these states, described in terms of vortex/antivortex pairs, are occupied,
organized, and affect each other energetically. An applied phase difference
over a large distance is shielded through excitation or rearrangement of
dressed vortex/antivortex pairs whose energy goes to zero with 
increasing pair size. Consequently, true glass correlations are expected only 
at $T=0$, which behaves as a critical point.

The rapid increase of the glass correlation length $\xi_{\rm th}\sim T^{-\nu}$ and the
spin-glass susceptibility $\chi_{\rm SG}(T)\sim T^{-\gamma}$,
with $\nu=1/|\theta|\simeq 2.2$ and $\gamma=(2-\eta)/|\theta|\simeq 4.4$ 
(note that $\eta\rightarrow 0$ as $T\rightarrow 0$), implies
glass-like behavior for a finite system at sufficiently low temperatures. 
Glass rigidity in this case arises from the freezing of vortex configurations
on the scale of the system size. In addition to the energy scale $\Delta E_{\rm pair}(R)$
for pair excitation on scale $R$, there should be another energy scale that
defines the energy barrier for vortex movement through continuous phase
deformation of the spins. Given that the effective disorder potential
$\tilde V({\bf r})$ seen by a single vortex has bound variations\cite{tang05},
this energy scale is of the order of the bare coupling constant $J$
at $T=0$ and could be reduced to smaller values by thermal excitations at finite temperatures.
This is a small scale phenomenon which has not been carefully analyzed so far
and may contribute to the apparent scaling under the assumption of a finite transition 
temperature $T_c\simeq 0.2J$ observed in some numerical studies.\cite{CP99,Kim,HMK}

Large scale vortex movement is essential for understanding dynamic relaxation
associated with phase slips. It also holds the key to interpreting
current-voltage characteristics when 
Eq. (\ref{gauge-glass-H}) is used to describe a superconducting film or 
Josephson-junction (JJ) array under suitable dynamic rules.\cite{newrock,tang03}
Due to the boundedness of $\tilde V({\bf r})$ 
in the ground state, a single vortex in the system is not localized
at any temperature $T>0$. If we ignore the screening effects
and take $\tilde V({\bf r})$ to be the actual potential when the vortex moves around in
the system, a finite diffusion constant 
\begin{equation}
D(T)\sim\langle e^{\tilde V({\bf r})/kT} \rangle^{-1}
\label{single-vortex-diff-const}
\end{equation}
can be assigned to the vortex.\cite{haus}
Due to spatial variations in $\tilde V({\bf r})$ and also the ``ageing'' type
relaxation through local rearrangement of the background vortices
when the vortex in question moves to a new site, the actual diffusion
constant may have a more complex temperature dependence than a
simple Arrhenius behavior with a single energy scale. Faster than
linear increase of the voltage with respect to an applied current $I$,
which drives vortices and antivortices in opposite directions,
may be expected at intermediate values of $I$.\cite{BJKim,chen08}
However, since the bottleneck for vortex movement resides at
small length scales in the present case, at a given $T$, linear I-V curve is expected
in the small current regime. Unlike the behavior in the ordered phase of the unfrustrated
JJ array\cite{tang03}, the size of this linear regime should not shrink significantly
when the system size increases. We leave a more detailed discussion of the
vortex dynamics in the gauge glass model to future publications.

\section*{Acknowledgements}
This work was partially supported by the Research Grants Council of the Hong Kong SAR
under grant 2020/04P, and by the Yukawa International Program for Quark-Hadron 
Sciences (YIPQS).

%

\end{document}